\documentclass[aps,prl,twocolumn,amsmath,amssymb,superscriptaddress,scrartcl]{revtex4-1}
\usepackage{extarrows}
\usepackage{slashed}
\usepackage{amsmath,amsfonts,amssymb,graphics,graphicx,epsfig,color,times,indentfirst,layout}
\usepackage{lipsum}
\usepackage{bbold}
\usepackage{extarrows}
\usepackage{slashed}
\usepackage{amsmath,amsfonts,amssymb,graphics,graphicx,epsfig,color,times,indentfirst,layout}
\usepackage{lipsum}
\usepackage{mathrsfs}
\usepackage[unicode=true,pdfusetitle,bookmarks=false,colorlinks=true,citecolor=black,urlcolor=black,linkcolor=black]{hyperref}
\usepackage{tikz}
\usepackage{tikz-cd}
\usetikzlibrary{arrows}
\usetikzlibrary{intersections}
\usetikzlibrary{shapes.geometric}

\begin{document}

\title{
Bridging global and local quantum quenches in conformal field theories
}

\date{\today}

\author{Xueda Wen}

\affiliation{
Institute for Condensed Matter Theory and Department of Physics,  University of
Illinois at Urbana-Champaign, 1110 West Green St, Urbana IL 61801, USA}

\affiliation{Kavli Institute for Theoretical Physics, University of California,
Santa Barbara, CA 93106, USA}

\begin{abstract}
Entanglement evolutions after a global quantum quench and a local quantum quench in 1+1
dimensional conformal field theories (CFTs)
show qualitatively different behaviors, and are studied within two different setups.
In this work, we bridge global and local quantum quenches in (1+1)-d CFTs in the same setup,
by studying the entanglement evolution from a specific inhomogeneous initial state.
By utilizing conformal mappings,
this inhomogeneous quantum quench is analytically solvable.
It is found that the entanglement evolution shows a global quantum quench feature in the short time limit,
and a local quantum quench feature in the long time limit.
The same features are observed in single-point correlation functions of primary fields.
We provide a clear physical picture for the underlying reason.
\end{abstract}
\maketitle


Global and local quantum quenches in (1+1) dimensional conformal field theories (CFTs) have received extensive
interest in both condensed matter physics and high energy physics recently \cite{Cardy1603,Cardy0503,Cardy0708,Hartman1303,Nozaki1302}.
In particular, the time evolution of entanglement entropy after a global/local quantum quench in (1+1)-d CFTs shows
universal features. Consider a semi-infinite subsystem $A=[0,\infty)$ on an infinite line.
In the case of a global quench, starting from an initial state which is translation invariant and short-range entangled,
the time evolution of entanglement entropy for $A$ grows linearly in $t$, \textit{i.e.}, $S_A(t)\sim t$ \cite{Cardy1603,Cardy0503}.
In the case of a local quench, two CFTs are connected at their ends ($x=0$) at time $t=0$ suddenly,
then one has $S_A(t)\sim \log t$ \cite{Cardy0708}.

\begin{figure}[ttt]
\includegraphics[width=3.50in]{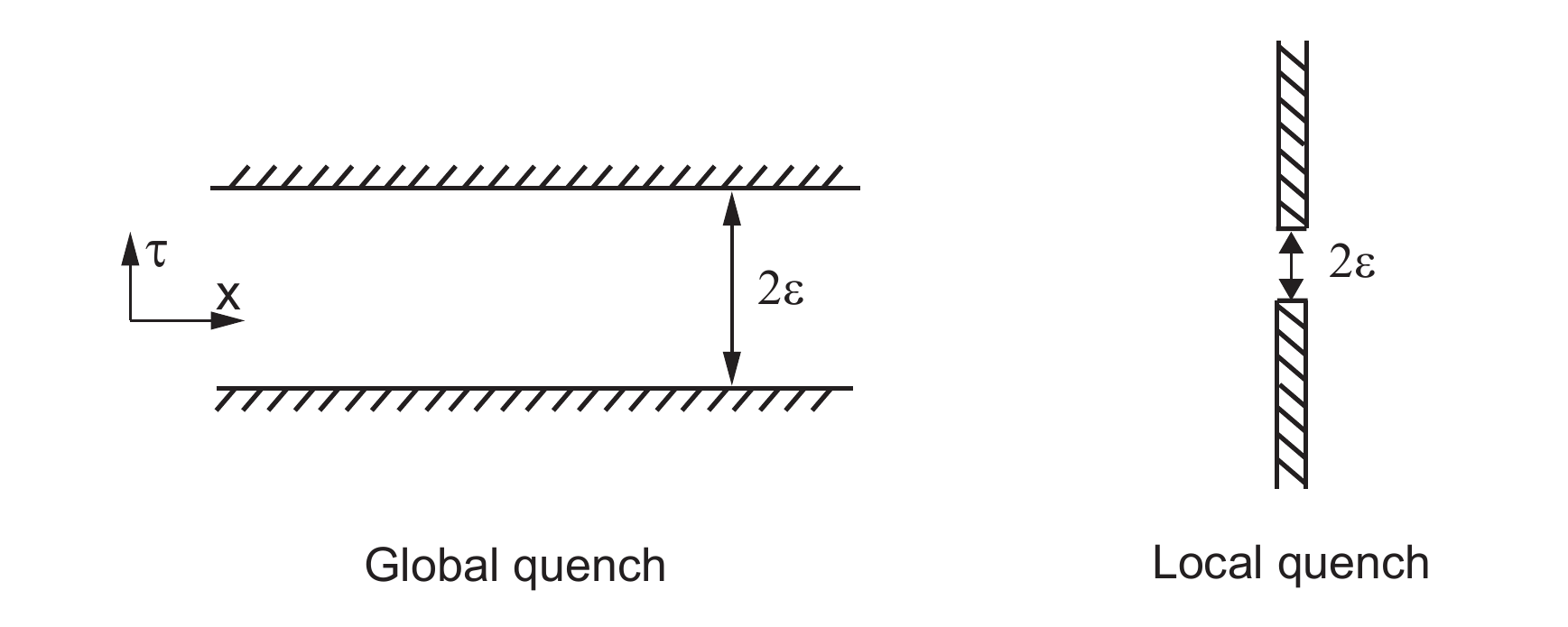}
\caption{Setups for global and local quantum quenches in (1+1)-d conformal field theories.
For the global case, one has a ``straight strip" with width $2\epsilon$. For the local case,
one has two slits located along the imaginary axis in Euclidean space.
}\label{conformalMap1a}
\end{figure}

Besides the different behaviors in $S_A(t)$,
the setups for studying global and local quantum quenches
in (1+1)-d CFTs are quite different. For the global case,
one may choose the initial state as \cite{Cardy1603,Cardy0503}
\begin{equation}\label{GlobalState}
|\psi_0\rangle=e^{-\epsilon H_{\text{CFT}}}|B\rangle,
\end{equation}
where $|B\rangle$ is a conformally invariant boundary state,
$H_{\text{CFT}}$ is the CFT Hamiltonian, and $\epsilon$
characterizes the correlation length (inverse mass).
For a translation invariant initial state, $\epsilon$ needs to be a constant,
and therefore is \textit{position independent}.
To calculate the entanglement entropy,
we are interested in the equal time correlation function of a set of
local operators $\mathcal{T}_i(x_i,t)$ which is called ``twist operator", within
the time evolved state $e^{-iH_{\text{CFT}}t}|\psi_0\rangle$.
In Euclidean space, this turns out to
be the calculation of correlation functions in a ``straight strip" with width $2\epsilon$,
as schematically shown in Fig.\ref{conformalMap1a}.
The boundary conditions at $\tau=\pm \epsilon$
are the conformally invariant boundary state $|B\rangle$.
Based on this setup, one can find the time
evolution of entanglement entropy for $A=[0, \infty)$ after a global quench as
\begin{equation}\label{GlobalQ}
S_A(t)=\frac{\pi c}{12\epsilon}t,
\end{equation}
where $c$ is the central charge of the underlying CFT. This
behavior can be well interpreted based on the quasi-particle picture \cite{Cardy1603,Cardy0503}.

For a local quantum quench, among different setups \cite{Cardy1603,Cardy0708,Nozaki1302,Nozaki1401,He1403}, we consider the case of connecting two
CFTs at their ends ($x=0$) at $t=0$.
In other words, for $t<0$, the two CFTs are decoupled and can not talk to each other.
Apparently, the initial state in this case is no longer translation invariant in space direction.
In Euclidean space, one needs to consider two slits
along the imaginary time axis, which are located at $(-i\infty,-i\epsilon)$ and
$(i\epsilon, i\infty)$, respectively (see Fig.\ref{conformalMap1a}).
The entanglement evolution for
$A=[0,\infty)$ after such a local quantum quench is \cite{Cardy0708}
\begin{equation}\label{localQ}
S_A(t)=\frac{c}{3}\log t.
\end{equation}

In this work, by studying a quantum quench from a specific inhomogeneous initial state,
we unify global and local quantum quenches in the same setup, but at different time scales.
It is found that the entanglement entropy evolution for $A=[0,\infty)$ is
\begin{equation}\label{Result0}
S_A(t)=
\left\{
\begin{split}
&\frac{\pi c}{12\epsilon} t,\quad\quad\quad\quad t\ll \Lambda,\\
&\frac{c}{3}\cdot\frac{\pi \Lambda}{4\epsilon}\cdot \log t, \quad t\gg \Lambda,
\end{split}
\right.
\end{equation}
where $\Lambda$ is a length scale we will introduce in the inhomogeneous initial state shortly. We also consider the case
$A=[l,\infty)$ with $l\gg \Lambda$, which shows very interesting features as well.

 \begin{center}
\textbf{Setup}
\end{center}

The inhomogeneous initial state we consider has the following expression
\begin{equation}\label{OurState}
|\psi_0\rangle=e^{-\epsilon(x)H_{\text{CFT}}}|B\rangle.
\end{equation}
Compared to the initial state in Eq.(\ref{GlobalState}), now $\epsilon(x)$ is
a function of position $x$. Physically, it means the correlation length is position dependent.
We choose $\epsilon(x)$ in the following way
\begin{equation}\label{hyperbolicBC}
\epsilon(x)=\sin\epsilon_0\cdot \sqrt{\Lambda^2+\left(\frac{x}{\cos\epsilon_0}\right)^2},
\end{equation}
where $\Lambda$ is a length scale. In particular,
we are interested in the limit $\epsilon_0\to 0$, and therefore
\begin{equation}
\epsilon(x)\simeq\epsilon_0 \cdot \sqrt{\Lambda^2+x^2}
\simeq\left\{
\begin{split}
&\epsilon,\quad\quad |x|\ll \Lambda,\\
&\frac{\epsilon |x|}{\Lambda}, \quad |x|\gg \Lambda,
\end{split}
\right.
\end{equation}
where, for comparison with the known results in global/local quenches, we have defined $\epsilon_0\Lambda=:\epsilon$.
Here $\epsilon$ is the parameter that appears in Eq.(\ref{GlobalState}).
Apparently, $\Lambda$ sets a length scale below which the initial state in Eq.(\ref{OurState})
looks the same as that in Eq.(\ref{GlobalState}).

\begin{figure}[t]
\includegraphics[width=2.50in]{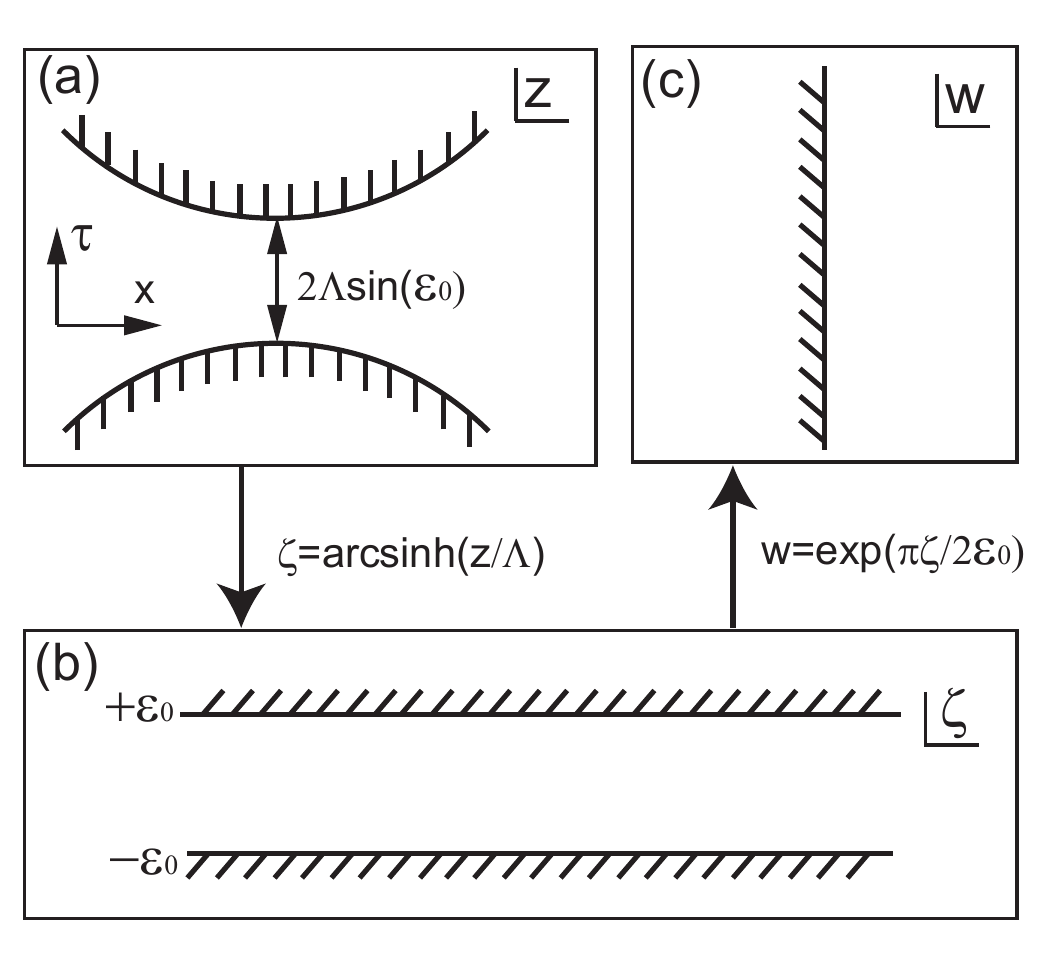}
\caption{Two-step conformal mapping in our calculation.
The ``hyperbolic strip"  is mapped to a ``straight strip" first, and then to the right half plane.
}\label{conformalMap3a}
\end{figure}

In Schr$\ddot{o}$dinger's picture, the time evolved state is
$|\psi(t)\rangle=e^{-iHt}|\psi_0\rangle$, where we write $H_{\text{CFT}}$ as $H$ for brevity. We are interested in the equal time correlation function of
a set of local operators $O_i(x_i)$, i.e.,
\begin{equation}
\begin{split}
&\langle \psi_0|e^{iHt}O_1(x_1)\cdots O_n(x_n)e^{-iHt}|\psi_0\rangle\\
=&\langle B|e^{iHt-\epsilon(x)H}O_1(x_1)\cdots O_n(x_n)e^{-iHt-\epsilon(x) H}|B\rangle.\nonumber
\end{split}
\end{equation}
In Euclidean space, this is nothing but the correlation function of local operators $O_i(z_i)$
in a ``hyperbolic strip'', with $z_i=x_i+i\tau$. The two edges of this
``hyperbolic strip" is defined by $\pm\epsilon(x)$ in Eq.(\ref{hyperbolicBC}), as schematically shown
in Fig.\ref{conformalMap3a} (a). On each edge, we have the boundary condition $|B\rangle$.
To calculate the correlation function in the ``hyperbolic strip", we use a two-step conformal mapping:
\begin{equation}\label{ConformalMapping}
\left\{
\begin{split}
\zeta=&\sinh^{-1}\left(\frac{z}{\Lambda}\right),\\
w=&\exp\left(\frac{\pi \zeta}{2\epsilon_0}\right),
\end{split}
\right.
\end{equation}
based on which we map the ``hyperbolic strip" in $z$-plane to a ``straight strip" in $\zeta$-plane,
 and then to the right half-plane in $w$-plane (see Fig.\ref{conformalMap3a}).
The problem is now reduced to the calculation of correlation functions in the right half-plane, which
is easier to handle.
To recover the real time evolution, we will analytically continue the imaginary times as $\tau\to it$ in the final step.

 \begin{center}
\textbf{Entanglement evolution for $A=[0,\infty)$}
\end{center}

\begin{figure}[ttt]
\includegraphics[width=2.05in]{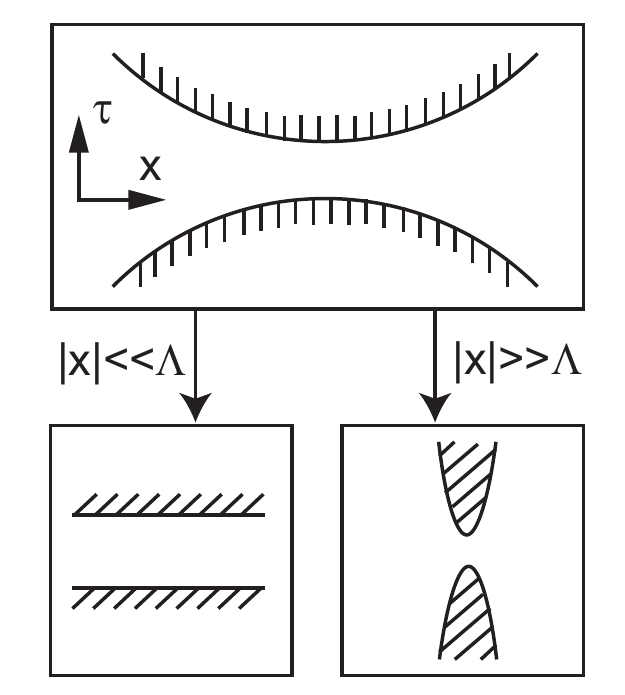}
\caption{How the ``hyperbolic strip" looks like in the short length scale
$|x|\ll \Lambda$ and long length scale $|x|\gg \Lambda$.}\label{PhysicsPicture}
\end{figure}

For simplicity, first we consider the case $A=[0,\infty)$.
The entanglement entropy of subsystem $A$ can be obtained by using the replica trick. One may first consider the $n$-th
Renyi entropy
\begin{equation}
S_A^{(n)}:=\frac{1}{1-n}\log\left[\text{tr}\left(\rho_A^n\right)\right].\nonumber
\end{equation}
And the von Neumann entanglement entropy is obtained by
\begin{equation}
S_A:=-\text{tr}\rho_A\log\rho_A=\lim_{n\to1}S_A^{(n)}.
\end{equation}
The term $\text{tr}(\rho_A^n)$ in $S^{(n)}_n$ is related with the correlation function of
$n$-th order twist operator $\mathcal{T}_n(z_0)$ as follows
\begin{equation}\label{DensityTwist}
\text{tr}\left(\rho_A^n\right)=\langle \mathcal{T}_n(z_0)\rangle_{\text{HYP}},
\end{equation}
where ``HYP" represents the ``hyperbolic strip", and $z_0=i\tau$.
The twist operator $\mathcal{T}_n(z)$ is a primary operator with the scaling dimension
\begin{equation}
x_n=\frac{c}{12}\left(n-\frac{1}{n}\right).\nonumber
\end{equation}
By using the conformal mapping in Eq.(\ref{ConformalMapping}), one has
\begin{equation}\label{twistOP}
\begin{split}
\langle \mathcal{T}_n(z_0)\rangle_{\text{HYP}}\sim&\left(\left|\frac{dw}{dz}\right|_{z=z_0}\frac{1}{2\text{Re}(w_0)}\right)^{x_n},\\
\end{split}
\end{equation}
which can be further simplified as
\begin{equation}\label{twistOP02}
\begin{split}
\langle \mathcal{T}_n(z_0)\rangle_{\text{HYP}}
\sim&\left(
\sqrt{1-\tau^2/\Lambda^2}\cdot2\cos\left(\pi\alpha/2\epsilon_0\right)
\right)^{-x_n}.
\end{split}
\end{equation}
Here we use ``$\sim$" instead of ``$=$" because we are mainly interested in the universal leading term, and the non-universal
coefficient will only contribute to non-universal constant terms in the entanglement entropy, which we are not interested in here.
$\alpha$ in Eq.(\ref{twistOP02}) is defined through
\begin{equation}
\exp\left(i\alpha\right):=i\frac{\tau}{\Lambda}+\sqrt{1-\frac{\tau^2}{\Lambda^2}}.\nonumber
\end{equation}
Next, by considering the analytical continuation $\tau\to it$, and noting that $\frac{\pi}{2\epsilon_0}\gg 1$, one can obtain
the analytical expression for $S_A(t)$ as follows:
\begin{equation}
S_A(t)=\frac{\pi c}{12}\cdot \frac{\Lambda}{\epsilon}\cdot \log \left(
\sqrt{1+\frac{t^2}{\Lambda^2}}+\frac{t}{\Lambda}
\right).
\end{equation}
This result holds for arbitrary time $t\gg \epsilon$. In the short time limit ($\epsilon\ll t\ll \Lambda$) and long time
limit ($t\gg \Lambda$), it reduces to the result in Eq.(\ref{Result0}).
One can find that in the short time limit, $S_A(t)$ is exactly the same as the global quantum quench result in Eq.(\ref{GlobalQ}).
The reason is as follows. In short time limit, based on the quasi-particle picture, only the region $|x|\sim t\ll \Lambda$
is relevant for the time evolution of $S_A(t)$. As shown in Fig.\ref{PhysicsPicture}, the ``hyperbolic strip" looks the
same as a ``straight strip" for $|x|\ll \Lambda$. In addition, the width of this ``straight strip" is
$2\Lambda\sin\epsilon_0\simeq 2\Lambda\epsilon_0=2\epsilon$ by definition, which is the same as
that in Fig.\ref{conformalMap1a}.

On the other hand, in the long time limit $t\gg \Lambda$, the ``hyperbolic strip" in the long-length scale $|x|\sim t$ looks like
``two slit" (see Fig.\ref{PhysicsPicture}), similar with the local quench setup in Fig.\ref{conformalMap1a}.
This is why we have $S_A(t)\sim \log t$ in the long time
limit. By comparing with the local quench result in Eq.(\ref{localQ}), it is found that
$S_A(t\gg \Lambda)$ in Eq.(\ref{Result0}) is enhanced by a factor $\pi\Lambda/4\epsilon$.
This is as expected by considering that the standard local quantum quench happens at one point (see Fig.\ref{conformalMap1a}),
while in Fig.\ref{PhysicsPicture} the ``local quench" happens in a region with the length scale
$\sim\Lambda$. Therefore, compared with the local quench setup in Fig.\ref{conformalMap1a},
more quasiparticles are emitted from the region $|x|\sim\Lambda$ and contribute to $S_A(t)$,
resulting in an enhancing factor $\pi\Lambda/4\epsilon$.

It is straightforward to check that the result in Eq.(\ref{Result0}) also holds for
$A=[l,\infty)$ with $|l|\ll \Lambda$. It is because in the short time limit $t\ll \Lambda$, the relevant physics
still happens in the ``straight strip", which corresponds to the global quench setup.
In the long time limit $t\gg \Lambda$ (and therefore a long length scale $|x|\gg \Lambda$),
the system cannot see the difference between $l=0$ and $|l|\ll \Lambda$.

\begin{center}
\textbf{Entanglement evolution for $A=[l,\infty)$ with $l\gg \Lambda$}
\end{center}

Based on the analysis above, it is expected that if the subsystem $A$ is far from the origin $x=0$,
one should observe a local quantum quench feature only. If this is true, another natural question
is: Is $S_A(t)$ enhanced by the same factor $\pi\Lambda/4\epsilon$?

Before checking these questions, let us remind ourselves of the result of local quench for
$S_A(t)$ with $A=[l,\infty)$, when connecting two CFTs at $x=0$:
\begin{equation}\label{Local2}
S_A(t)=
\left\{
\begin{split}
&\frac{c}{6}\log l,\quad\quad\quad \quad t<l,\\
&\frac{c}{6}\log (t^2-l^2), \quad t>l.
\end{split}
\right.
\end{equation}
The fact that $S_A(t)$ is time independent for $t<l$ simply means that the
quasiparticles emitted from $x=0$ have not arrived at subsystem $A$.
In addition, in the long time limit $t\gg l$, $S_A(t)$ is reduced to Eq.(\ref{localQ}),
as expected.

Now let us check the behavior of $S_A(t)$ for $A=[l,\infty)$ in our setup.
To capture the local quench feature, we consider the limit $l\gg \Lambda$.
The calculation of $\text{tr}\left(\rho_A^n\right)$ is the same as Eq.(\ref{DensityTwist}),
but now we use $z_0=l+i\tau$.
Following similar procedures in the case of $A=[0,\infty)$, one can obtain
\begin{equation}\label{SingleT}
\langle \mathcal{T}_n(z_0)\rangle_{\text{HYP}}\sim\left(\frac{\rho}{\Lambda}\cdot 2\cos(\pi\alpha/2\epsilon_0)\right)^{-x_n},
\end{equation}
where $\alpha$ is defined by
\begin{equation}\label{BigLalpha}
\exp\left(i\alpha\right):=\frac{l+\rho\cos\theta+i\left(\tau+\rho\sin\theta\right)}{\sqrt{(l+\rho\cos\theta)^2+(\tau+\rho\sin\theta)^2}},
\end{equation}
with
\begin{equation}\label{rhoTheta}
\left\{
\begin{split}
\rho&=\left[(\Lambda^2+l^2-\tau^2)^2+4l^2\tau^2\right]^{1/4},\\
\theta&=\frac{1}{2}\arctan\frac{2 l\tau}{\Lambda^2+l^2-\tau^2}.
\end{split}
\right.
\end{equation}
Then we do analytical continuation $\tau\to it$. As shown in the appendix, the analytical continuation
depends on whether $t<l$ or $t>l$ \cite{Cardy0708,Wen1501}. After some straightforward algebra, one can obtain the leading order
of $S_A(t)$ as
\begin{equation}\label{BigL}
S_A(t)=
\left\{
\begin{split}
&\frac{c}{6} \log l,\quad\quad\quad\quad\quad\quad t\ll l,\\
&\frac{c}{6}\cdot \frac{\pi\Lambda}{4\epsilon}\cdot \log (t^2-l^2), \quad l>t.
\end{split}
\right.
\end{equation}
In addition, for $t\ll l$, if we check the leading order in time $t$, its contribution to
$S_A(t)$ is
\begin{equation}\label{subleadingT}
S_A '(t)=\frac{\pi c}{12\epsilon}\cdot \frac{\Lambda}{l}\cdot t, \quad t\ll l.
\end{equation}
Several remarks are in order:

-- By comparing Eq.(\ref{BigL}) with Eq.(\ref{Local2}),
$S_A(t)$ in Eq.(\ref{BigL}) indeed shows the feature of a local quantum quench.
In particular, for $t\ll l$, $S_A(t)$ is exactly the same as that in local quench, simply because the
regions $[0,\infty)$ and $(-\infty,0]$ are effectively decoupled before $t=0$.
For $t>l$, compared with the local quench result in Eq.(\ref{Local2}), there is an enhancing factor
$\pi\Lambda/4\epsilon$, which has the same physical origin as that in Eq.(\ref{Result0}).
This can be viewed as a consistent double check. For $t\gg l$, $S_A(t)$ is reduced to
$S_A(t)=\frac{c}{3}\cdot \frac{\pi \Lambda}{4\epsilon}\cdot \log t$, as expected.

-- Although the leading order of $S_A(t)$ shows the local quench feature, if we look at the
time dependent term of $S_A(t)$ for $t\ll l$ [see Eq.(\ref{subleadingT})], it grows linearly in $t$,
which is the feature of a global quantum quench. Compared with Eq.(\ref{GlobalQ}), the growth rate
is suppressed by a factor $\Lambda/l$. The reason is as follows.
For $t\ll l$, only the degrees of freedom near the entanglement
surface ($x\sim l$) contribute to $S_A(t)$. If we look at the ``hyperbolic strip" locally around $x\sim l$,
it is a ``straight strip" with a width $2\epsilon\cdot \frac{l}{\Lambda}$.
This explains the factor $\Lambda/l$ in Eq.(\ref{subleadingT}).

As a short sum, for the time dependent terms in $S_A(t)$ with $A=[l,\infty)$, we can observe a
global quantum quench feature in the short time limit, and a local quantum quench feature
in the long time limit, for both $l\ll \Lambda$ and $l\gg \Lambda$.

\begin{center}
\textbf{One-point function of primary fields}
\end{center}

It is also interesting to check the time evolution of
single point correlation function of a generic primary field $\Phi(x,t)$.
The expressiones of $\langle \Phi(x,t)\rangle$ for standard global and local quantum quenches
in Fig.\ref{conformalMap1a} are as follows. For a global quantum quench, one has \cite{Cardy1603}:
\begin{equation}
\langle \Phi(x,t)\rangle\sim \exp\left(-\frac{\pi \Delta_{\Phi}}{2\epsilon}t\right),
\end{equation}
where $\Delta_{\Phi}$ is the scaling dimension of the primary field $\Phi(x,t)$.
For a local quantum quench, one has \cite{Cardy0708}
\begin{equation}
\langle \Phi(x,t)\rangle\sim
\left\{
\begin{split}
&x^{-\Delta_{\Phi}}, \quad \quad \quad \quad\quad t< x,\\
&\left(t^2-x^2\right)^{-\Delta_{\Phi}}, \quad t>x.
\end{split}
\right.
\end{equation}
For $t\gg x$, one has $\langle \Phi(x,t)\rangle\sim t^{-2\Delta_{\Phi}}$.
Now we turn to the inhomogeneous quantum quench in our setup.
The calculation of $\langle \Phi(x,t)\rangle$ is the same as evaluating
$\langle \mathcal{T}_n(z_0)\rangle_{\text{HYP}}$
in Eqs.(\ref{twistOP}) and (\ref{SingleT}). For $x\ll \Lambda$, it is found that
\begin{equation}
\langle \Phi(x,t)\rangle\sim
\left\{
\begin{split}
&\exp\left(-\frac{\pi \Delta_{\Phi}}{2\epsilon}t\right), \quad t\ll \Lambda,\\
&t^{-\frac{\pi\Lambda}{4\epsilon}\cdot 2\Delta_{\Phi}}, \quad\quad\quad\quad t\gg \Lambda.
\end{split}
\right.
\end{equation}
One can observe that $\langle \Phi(x,t)\rangle$ shows an exponential decay for $t\ll \Lambda$, and
a power law decay for $t\gg\Lambda$. In particular, the form of $\langle \Phi(x,t\ll\Lambda)\rangle$
is exactly the same as that for a global quench. For $t\gg\Lambda$,
$\langle \Phi(x,t)\rangle$ shows the same behavior as a local quantum quench, but with
a factor $\pi\Lambda/4\epsilon$ difference, which
has the same origin as that in Eqs.(\ref{Result0}) and (\ref{BigL}).

For $x\gg \Lambda$, one has
\begin{equation}
\langle \Phi(x,t)\rangle\sim
\left\{
\begin{split}
&x^{-\Delta_{\Phi}}\cdot\exp\left(-\frac{\pi \Delta_{\Phi}}{2\epsilon}\cdot\frac{\Lambda}{x}\cdot t\right), \quad t\ll x,\\
&\left(t^2-x^2\right)^{-\frac{\pi\Lambda}{4\epsilon}\cdot \Delta_{\Phi}}, \quad\quad\quad\quad\quad\quad t>x.
\end{split}
\right.
\end{equation}
In the short time limit $t\ll x$,
the first term $x^{-\Delta_{\Phi}}$, which is time independent, is exactly the same as that for a local quantum
quench.
The second term shows an exponential decay in time, corresponding to the feature of a global quantum quench.
But the decay rate is now suppressed by a factor $\Lambda/x$, which has the same physical origin as
the factor $\Lambda/l$ in Eq. (\ref{subleadingT}).
For $t>x$, $\langle \Phi(x,t)\rangle$ has the same expression
as that for a local quench, but with a factor $\pi\Lambda/4\epsilon$ difference.
Again, this factor has the same origin as that in Eqs.(\ref{Result0}) and (\ref{BigL}).
All the features above are consistent with our analysis on entanglement evolution.

 \begin{center}
\textbf{Discussion and Conclusion}
\end{center}


In this work, we propose a specific inhomogeneous quantum quench,  which is analytically solvable,
to bridge the global and local quantum quenches that were previously studied
in two totally different setups in (1+1)-d CFTs.
Given the initial state in Eqs.(\ref{OurState}) and (\ref{hyperbolicBC}), we study the entanglement evolution
$S_A(t)$ for $A=[l,\infty]$ with $l\ll \Lambda$ and $l\gg\Lambda$, respectively.
For $l\ll \Lambda$, it is found that $S_A(t)$ shows a global quantum quench feature in the short time limit, and
a local quantum quench feature in the long time limit.
For $l\gg \Lambda$, the leading term of $S_A(t)$ shows a local quantum quench feature in both short and long time limit.
However, for the time dependent term in short time limit, it shows the global quantum quench feature.
Therefore, we claim that for the time dependent terms in $S_A(t)$ with $A=[l,\infty)$,
one can observe the global quantum quench feature
in short time limit, and the local quantum quench feature in long time limit,
no matter $l\ll \Lambda$ or $l\gg\Lambda$.
We also check the time evolution of single-point correlation functions of primary fields, the behavior of which
is consistent with the entanglement evolution.

There are many other interesting setups for inhomogeneous quantum quenches \cite{Cardy0808,Calabrese0804,Allegra2016,Dubail1606},
among which the most relevant one to ours
is the work by Sotiriadis and Cardy \cite{Cardy0808}, although the motivation is different.
In Ref.\cite{Cardy0808}, the authors studied a quantum quench
from an inhomogeneous initial state, which can be mapped to a ``straight strip", with the same spirit as ours.
However, the specific configuration under study is a ``stepped width strip", and
only global quantum quench features are observed. We also noted some interesting setups in the holographic entanglement
context \cite{Tokiro0}. In Ref.\cite{Tokiro1}, various quantum operations including local projection measurements and
partial swapping in CFTs and their holographic dual are studied. Both global and local quantum quench features may arise
in the entanglement evolution after such quantum operations. It is our future aim to give a systematic study of different
setups that may give rise to global and local quantum quench features at different time scales.
A possible method is to start from a ``straight strip", and map it to a ``strange strip" by considering various interesting
conformal mappings. Then we choose this ``strange strip" as the inhomogeneous initial state for quantum quenches.
The calculation can be done by mapping it back to the ``straight strip".
This is, in spirit, similar to the methods used in Ref.\cite{Wen1604,Cardy1608}, although with different motivations.

Our setup may also be useful for studying related problems, e.g., the speed limit of entanglement propagation from
an initial state which is not translation invariant \cite{Hartman1512}. It is also interesting to study
the holographic dual of inhomogeneous quantum quenches based on our setup.

\begin{center}
\textbf{Acknowledgement}
\end{center}

XW thanks the ``Team quench" Journal Club at KITP,
the discussion during which stimulates his interest in studying this problem.
XW also thanks Tokiro Numasawa, Tomonori Ugajin, Shinsei Ryu, and Andreas W.W. Ludwig for helpful discussions.
This work is supported by KITP Graduate fellowship.

\begin{center}
\textbf{Appendix}
\end{center}

In this appendix, we give some details of calculating $S_A(t)$ for $A=[l,\infty)$ with $l\gg \Lambda$.
Starting from $\langle \mathcal{T}_n(z_0)\rangle_{\text{HYP}}$ in Eq.(\ref{SingleT}), we do analytical
continuation $\tau\to it$.
It is noted that $\rho\cos\theta$ and $\rho\sin\theta$
depend on whether $t<l$ or $t>l$ as follows: \cite{Cardy0708,Wen1501}
\begin{equation}\label{AnalyticalT}
\small{
\left\{
\begin{split}
\rho\cos\theta\to& \text{max}[l,t]\left(1+\frac{\Lambda^2}{2\left[(\text{max}[l,t])^2-(\text{min}[l,t])^2\right]}\right)\\
\rho\sin\theta\to& i\text{min}[l,t]\left(1+\frac{\Lambda^2}{2\left[(\text{min}[l,t])^2-(\text{max}[l,t])^2\right]}\right),
\end{split}
\right.}
\end{equation}
where $\rho$ and $\theta$ are defined in Eq.(\ref{rhoTheta}). We are interested in the limit $l,t\gg \Lambda$, so that
\begin{equation}
\rho\simeq\sqrt{|l^2-t^2|}.
\end{equation}
For $t< l$, it is found that Eq.(\ref{BigLalpha}) can be expressed as
\begin{equation}
\exp(i\alpha)\simeq \sqrt{\frac{l-t}{l+t}}.
\end{equation}
In the limit $\frac{\pi }{4\epsilon_0}\gg 1$, one has
\begin{equation}
\begin{split}
S_A^{(n)}(t)\simeq &\frac{c}{12}\cdot \frac{n+1}{n}
\log\left(\sqrt{l^2-t^2}\cdot \left(\frac{l+t}{l-t}\right)^{\frac{\pi }{4\epsilon_0}}\right).\\
\end{split}
\end{equation}
Noting that $\epsilon:=\epsilon_0\Lambda$, one can obtain Eqs.(\ref{BigL}) and (\ref{subleadingT})
in the short time limit $t\ll l$.
For $t>l$, based on Eq.(\ref{AnalyticalT}), one can obtain
\begin{equation}
\begin{split}
\exp\left(i\alpha\right)\simeq \frac{\Lambda}{2}\cdot \frac{1}{\sqrt{t^2-l^2}}.
\end{split}
\end{equation}
In the limit $\frac{\pi }{4\epsilon_0}\gg 1$, one has
\begin{equation}
\small{
\begin{split}
S_A^{(n)}(t)\simeq &\frac{c}{12}\cdot \frac{n+1}{n}
\log\left(\sqrt{t^2-l^2}\cdot \left(\frac{2\sqrt{t^2-l^2}}{\Lambda}\right)^{\frac{\pi }{2\epsilon_0}}\right),\\
\end{split}}
\end{equation}
based on which one can obtain the leading term in Eq.(\ref{BigL}) for $t>l$.

\end{document}